\documentclass[sigconf]{acmart}

\AtBeginDocument{%
  \providecommand\BibTeX{{%
    \normalfont B\kern-0.5em{\scshape i\kern-0.25em b}\kern-0.8em\TeX}}}

\acmConference[]{}{}{}

\copyrightyear{2020}
\acmYear{2020}
\setcopyright{rightsretained}

\usepackage{balance}
\usepackage{graphics}
\usepackage{booktabs}
\usepackage{subcaption}
\usepackage{textcomp}
\usepackage[htt]{hyphenat}
\usepackage{xcolor}
\usepackage{multirow}
\usepackage{gensymb}
\usepackage{xspace}
\usepackage{tabularx}

\usepackage{microtype}        
\usepackage{ccicons}          

\usepackage{todonotes}

\usepackage{hyperref}

\newcommand{\redacted}[1]{\textbf{REDACTED}}

\newcommand{\urllink}[1]{\footnote{\url{#1}}}

\newcommand{\figref}[1]{Figure~\ref{fig:#1}}
\newcommand{\tabref}[1]{Table~\ref{tab:#1}}

\newcommand*{\etal}{et~al.\xspace}
\newcommand*{\eg}{e.g.,\xspace}

\newcommand*{\projectname}{Project Calico\xspace}

\newenvironment{s_enumerate}{
\begin{enumerate}
  \setlength{\itemsep}{3pt}
  \setlength{\parskip}{0pt}
  \setlength{\parsep}{0pt}
}{\end{enumerate}}

\def\plaintitle{\projectname{}: Wearable Chemical Sensors for Environmental Monitoring}
\def\plainauthor{First Author, Second Author, Third Author,
  Fourth Author, Fifth Author, Sixth Author}

\def\plainkeywords{Chemical sensing; cosmetic computing; smart materials; environmental sensors; integrated wearables}

\definecolor{linkColor}{RGB}{6,125,233}
\hypersetup{%
  pdftitle={\plaintitle},
  pdfauthor={\plainauthor},
  pdfkeywords={\plainkeywords},
  pdfdisplaydoctitle=true,
  bookmarksnumbered,
  pdfstartview={FitH},
  colorlinks,
  citecolor=black,
  filecolor=black,
  linkcolor=black,
  urlcolor=linkColor,
  breaklinks=true,
  hypertexnames=false
}

\begin{document}

\title{\plaintitle}



\def\UW{\footnotemark[\large{$^\star$}]}
\def\MSR{\footnotemark[\large{$^\dagger$}]}
\def\DePaul{\footnotemark[\large{$^\ast$}]}

\author{Alex Mariakakis$^{*}$, Sifang Chen$^{*}$, Bichlien Nguyen$^\circ$, Kirsten Bray$^\dagger$, Molly Blank$^\circ$, Jonathan Lester$^\circ$, Lauren Ryan$^\circ$, Paul Johns$^\circ$, Gonzalo Ramos$^\circ$, Asta Roseway$^\circ$}
\affiliation{\vspace{0.15cm}\institution{$^{*}$University of Washington, Seattle, WA; $^\dagger$DePaul University, Chicago, IL; $^\circ$Microsoft Research, Redmond, WA}}
\email{{atm15, sifangc}@uw.edu; kbray4@depaul.edu; {bnguy, v-mobla, jonathan.lester, lryan, paul.johns, goramos, astar}@microsoft.com}



\renewcommand{\shortauthors}{Mariakakis et al.}

\begin{abstract}
Environmental hazards often go unnoticed because they are invisible to the naked eye, posing risks to our health over time. 
\projectname{} aims to raise awareness of these risks by augmenting everyday fashion with color-changing chemical sensors that can be observed at a glance or captured by a smartphone camera. 
\projectname{} leverages existing cosmetic and fabrication processes to democratize environmental sensing, enabling creators to make their own accessories.
We present two fashionable instantiations of \projectname{} involving UV irradiation. 
EcoHair, created by hair treatment, is UV-sensitive hair that intensifies in color saturation depending on the UV intensity.
EcoPatches, created by inkjet printing, can be worn as temporary tattoos that change their color to reflect cumulative UV exposure over time. 
We present findings from two focus groups regarding the \projectname{} vision and gathered insights from their overall impressions and projected use patterns.
\end{abstract}



\maketitle

\section{Introduction}
\label{sec:intro}

\begin{figure}[!t]
\begin{center}
\includegraphics[width=\columnwidth]{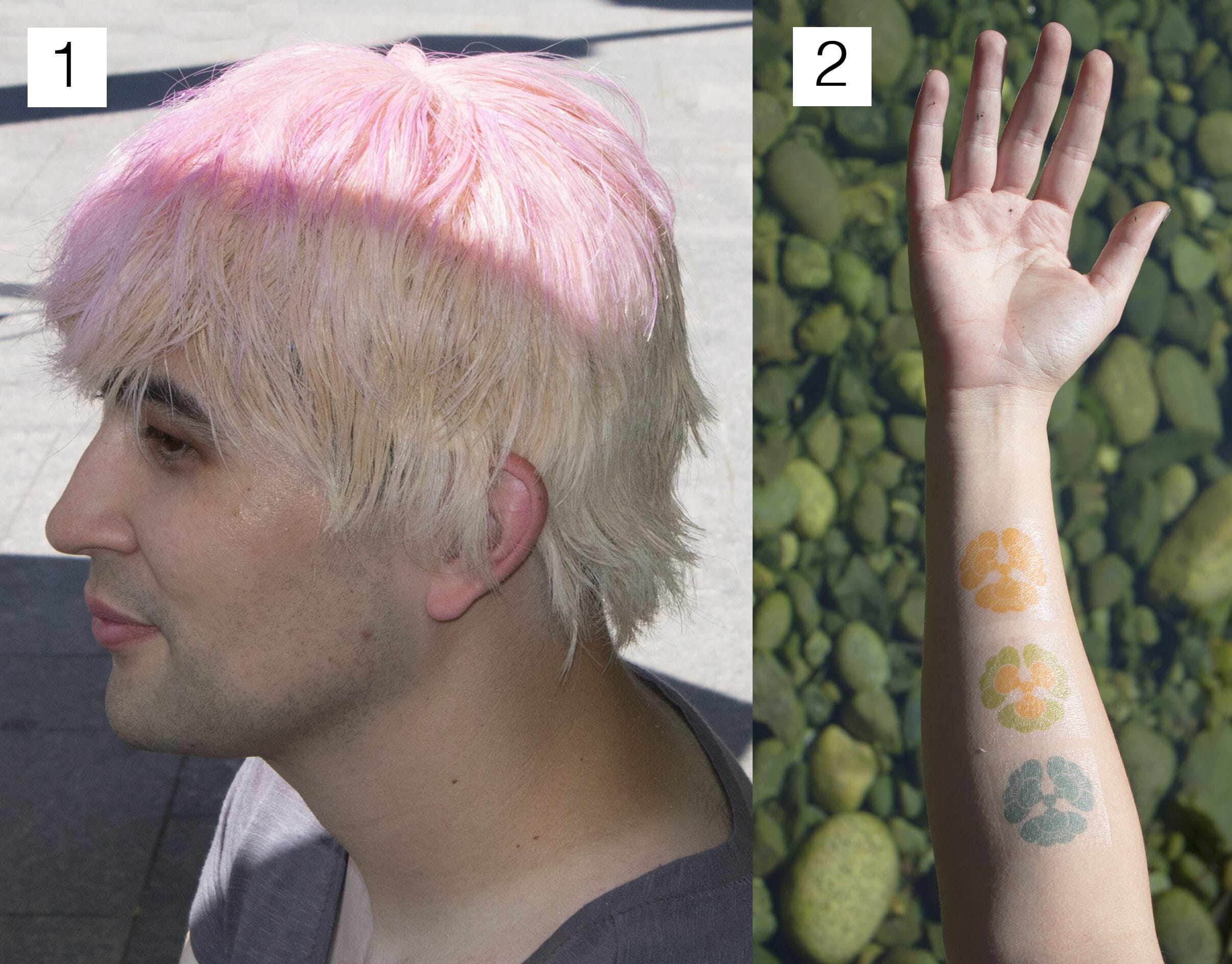}
\end{center}
\caption{\projectname{} aims to democratize and incentivize the measurement of environmental hazards using dynamic fashion. (1) EcoHair changes color to show instantaneous UV intensity. (2) EcoPatches show cumulative UV exposure.}
\label{fig:moneyshot}
\vspace{-0.2in}
\end{figure}

Environmental hazards can adversely affect our health in subtle ways because they are not easily perceived by individuals. As a result, their effects often go unnoticed until the hazard levels exceed dangerous thresholds and cause visible medical symptoms. Visualizing and quantifying these hazards are essential for raising public awareness and reducing incidences of exposure. Currently, environmental monitoring and reporting are mainly performed by environmental agencies using specialized instruments and sensor networks. These networked sensors are accurate and reliable but prohibitively expensive to distribute to individuals. Thus, conventional sensors fail to account for variations in exposure levels due to individual lifestyle differences \cite{Akiba1999,KUMAR2015199,Parkin2011}.

Crowdsourcing data from low-cost electronic sensors has been gaining momentum as a new paradigm for high-resolution environmental monitoring \cite{Muller2015,KamelBoulos2011,KUMAR2015199}. However, these devices do not integrate well with everyday life, which could explain why there has yet to be a notable deployment of them to this date. Past research has explored ways of integrating sensing with fashion \cite{Guler2016}, yet electronic accessories still require power and are too delicate to merge with traditional clothing. An alternative approach that has been much less explored in the UbiComp research community is to make these wearables directly out of responsive materials where sensing and actuation are performed by chemical computation. Despite the growing body of research on chemical sensors \cite{Araki2017,Kan2017,Kao2017EarthTones,Shi2018}, they have yet to become practical for everyday use.

To address these issues, \projectname{} (\figref{moneyshot}) introduces a collection of chemical-based sensors that serve as both personalized environmental hazard detectors and fashion accessories. The objectives of \projectname{} are three-fold: 
\begin{s_enumerate}
    \item To promote citizen science by making environmental monitoring more convenient,
    \item To generate interest in chemical sensing by leveraging existing cosmetic and fabrication processes, and
    \item To demonstrate that embedded chemical sensing can improve the aesthetics and functionality of everyday objects.
\end{s_enumerate}

\projectname{} accessories are designed to be easily interpretable without assistance. However, by transducing invisible environmental hazards into quantifiable visible color changes, \projectname{} accessories also present an opportunity for computer vision-based techniques to measure environmental hazards that normally require specialized hardware. In the near future, we envision a \projectname{} companion app that interprets a fashion accessory for the user, tracks measurements over time, and connects users with a wider network of \projectname{} citizen scientists.

In this paper, we focus on \projectname{} accessories that react to ultraviolet (UV) radiation. UV radiation is a major risk factor for sunburns and skin cancer. Skin cancer is the most commonly diagnosed cancer in the United States, yet the risk of skin cancer can be significantly reduced with proper sunscreen application and sunlight avoidance \cite{Armstrong2001,Siegel2011}. Furthermore, UV levels have been affected by complex interactions between climate change and ozone depletion \cite{Herman2010,McKenzie2011,Watanabe2011}; crowdsourced data from widely distributed UV sensors could be useful for inferring ozone depletion rates. 

We describe the design and manufacturing processes for two accessories: EcoHair and EcoPatches. EcoHair uses photochromic hair dyes that can be easily incorporated with standard hair products to display instantaneous UV intensity. EcoPatches are inkjet-printable stickers that display cumulative UV exposure with custom UV-sensitive inks. We conducted two focus groups to gauge user interest in our prototypes and elicit design recommendations for future \projectname{} sensors. Topics of discussion included aesthetics that incentivize exposure, designs that are aesthetically pleasing during transitions, and the permanence of aesthetic changes.
\section{Related Work}
\label{sec:related}

The growing interest in embedded sensors is exemplified by a recent surge in products and research prototypes involving materials that change their appearance in the presence of ambient phenomena. Below, we highlight work in colorimetric sensing, specifically for UV radiation, and fashion accessories that act as sensors.

\subsection{UV Sensing}
UV sensors can be divided into two categories: dosimeters for measuring cumulative UV exposure and radiometers for measuring instantaneous UV intensity. 
These devices often cost hundreds of dollars and are too bulky to be carried around during outdoor activities when UV exposure is most pronounced \cite{Banerjee2017ARadiometers, noh2001investigation}. There has been ongoing research and commercial efforts to reduce the cost and size of UV sensors so they can be more easily accessed by the average person. For example, L'Or\'{e}al recently presented a battery-free sensor designed to be worn on the thumbnail\urllink{http://www.lorealusa.com/media/press-releases/2018/january/uv-sense}.

A number of chemical-based UV sensors have also emerged in recent years. EarthTones is a conceptual line of novel cosmetic powders and pigments that exhibit UV-induced, irreversible color changes \cite{Kao2017EarthTones}. Because photosensitive chemicals may cause skin irritation, portable form factors that encapsulate these chemicals are preferable. Araki \etal{} \cite{Araki2017} created a temporary tattoo that changes color based on temperature changes or UV exposure. L'Or\'{e}al's My UV Patch\urllink{https://www.laroche-posay.us/my-uv-patch} and LogicInk\urllink{https://logicink.com/} are commercially available UV-sensing patches. Both patches integrate UV-sensitive chemicals in a polymer membrane for measuring instantaneous and cumulative UV exposure. Making these patches requires access to wet lab and cleanroom facilities, which makes prototyping expensive and difficult. In contrast, \projectname{} accessories can be made using processes that are already familiar to most people (\eg{} inkjet printing and hair treatment), making rapid prototyping more tenable for the average hobbyist. Furthermore, \projectname{} accessories are designed to be easily interpreted; the wearer should be able to glance at their accessory without the assistance of technology and infer a measurement value.

\subsection{Environmentally Reactive Fashion}
Vibrant colors and dramatic physical designs are used as mechanisms for self-expression, but designers have also started to explore ways for fashion to change appearance dynamically.
Changes can be manually actuated by switches \cite{MartinSmall2011}, driven by embedded electronics \cite{Poupyrev2016}, or facilitated by functional materials that respond to one's surroundings \cite{Yao2015}; the latter category is of the most interest to our work.

One common functional material property that has reached commercial application is thermochromism, wherein a color change is caused by shifting temperatures \cite{white1999thermochromism}.
Thermochromism has been used in items like mood rings, mugs, and childrens' toys.
More recently, H\"{a}irI\"{O} \cite{Dierk2018} activates changes in hair color and shape using heat-activated nitinol wire, while THE UNSEEN\urllink{https://theunseenessence.com/fire-hair/} uses thermochromic inks in synthetic wigs.
Dynamic aesthetics have also been driven by changes in pH.
Kan \etal{} \cite{Kan2017} use anthocyanin, vanillin, and chitosan to dope materials and enact changes in color, shape, and odor.
Kim \etal{} \cite{kim2016wearable} enclose pH indicators in a polyvinyl chloride matrix to create a wearable fingernail sensor.
In cosmetics, Red 27 is an ingredient that appears colorless when dissolved in a waterless base but turns bright pinkish-red when it comes into contact with moisture.
This enables color-changing properties in products like lip gloss\urllink{https://www.physiciansformula.com/ph-matchmaker-ph-powered-lip-gloss.html} and lipstick\urllink{https://www.moodmatcher.com/products.cfm?Page=Products&Product=MM\%20Moodmatcher\%20Lipstick\%20-\%20Group}.
As a final example of fashionable environmental sensing, Aerochromics\urllink{http://aerochromics.com/} produces shirts that display dynamic patterns when they detect high levels of air pollutants. 

The aforementioned examples are primarily driven by aesthetics and novelty.
Those incentives, combined with the convenience of fashion as a deployable medium, present a unique research opportunity for fashion to act as an ambient environmental sensor.
We explore programmable materials that not only change their appearance automatically, but also provide some quantification about the user's surrounding environment.
The chemical sensors under \projectname{} are designed to be interpretable without aid, but there is also room for computer vision-based analysis to supplement the experience.
\section{Design and Fabrication}
\label{sec:design}
In this section, we describe the fabrication processes and design considerations for making EcoHair and EcoPatches.

\subsection{UV Radiometer EcoHair} 
\subsubsection{Motivation}
The head is often the part of the body that is first exposed to UV radiation, making it a prime location for measuring UV exposure. Furthermore, having natural hair or wearing a wig is a completely passive activity that requires no cognitive burden. With this in mind, EcoHair conveys UV exposure through hair color modifications like hair dyeing, mousse application, or hair extensions. Most people are already familiar with these practices, which lowers the adoption barrier of chemical sensing.
EcoHair is designed to indicate instantaneous UV intensity. UV intensity is reported by a UV index (UVI). UVI 2 and below is  typical for cloudy days, while a UVI 6 and above warrants protection.

\subsubsection{Fabrication}

EcoHair dyes rely on pigments made up of photochromic chemical compounds that can switch between two molecular states with distinct absorption spectra. In essence, the pigments show one color in the absence of UV light and a different color in the presence of UV light. Because the number of radiation induced chemical reactions is proportional to the number of incident photons \cite{Stranius2017}, pigment colors will appear more saturated at high UVI. The response time, the time it takes for the photochromic pigments to achieve full saturation, decreases when the UV intensity increases. Additionally, the response time is also a function of the surrounding matrix's composition and the molecular structure of the pigment \cite{Vikova2011}. 


There is a myriad of commercially available pigments with different pre- and post-UV exposure colors\urllink{https://solarcolordust.com/}. To create EcoHair dye, the appropriate pigments are mixed with water in 1 teaspoon of pigment powder for every 5 mL of water. EcoHair mousse is made in a similar manner, with the ratio between mousse and pigment depending on the desired color intensity. Once the suspension is created, it is applied to the hair, combed for even distribution, and dried under high heat to remove the solvent. 


\subsubsection{Potential Designs}
\begin{figure}[!tb]
\begin{center}
\includegraphics[width=\columnwidth]{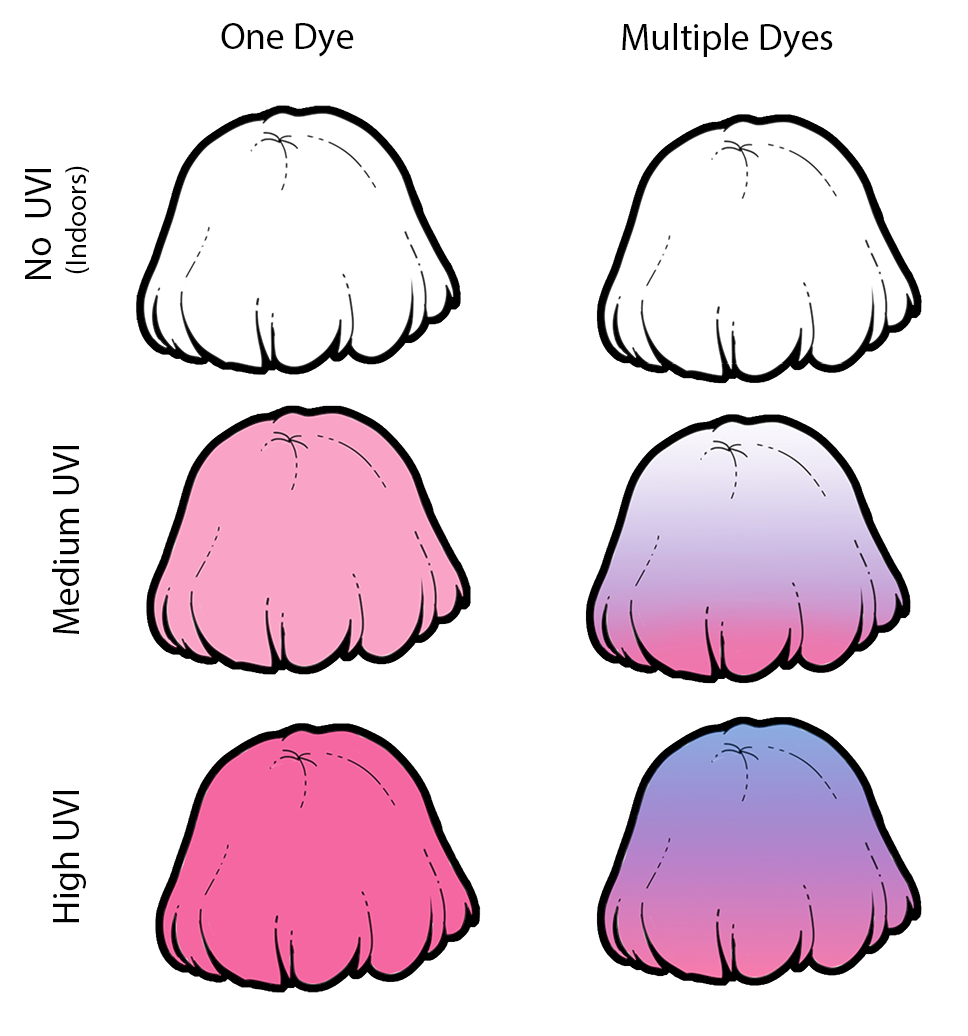}
\end{center}
\caption{We envision two different instantiations of EcoHair for conveying increasing levels of UV intensity: (left) a monochromatic design that changes its saturation, and (right) a tiered design that propagates saturation changes up the person's head.}
\label{fig:hair-design}
\end{figure}

\begin{figure}[!tb]
\begin{center}
\includegraphics[width=\columnwidth]{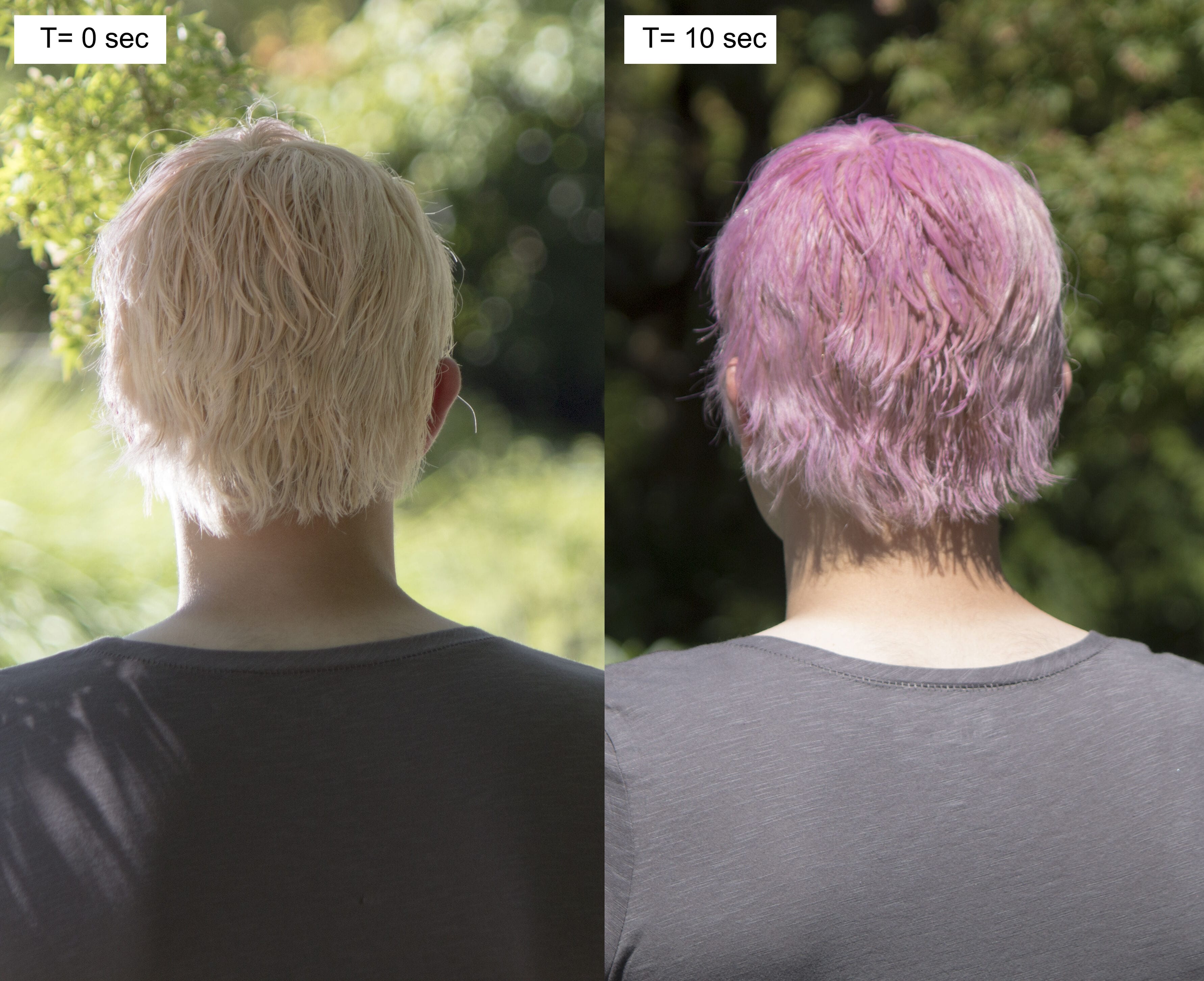}
\end{center}
\caption{A person with monochromatic EcoHair walking out of the shade during a day at UVI 7.}
\label{fig:hair-real}
\end{figure}

EcoHair can be applied on any type or color of hair; however, white- and platinum blonde-colored hair leads to the best color contrast. It is best for the initial colors of the pigments to match the hair's ordinary color so that it blends in naturally. \figref{hair-design} shows two different ways that EcoHair can convey instantaneous UV intensity in a quantitative manner. Along the left side of the figure, a single tone wig increases its saturation with increasing intensity up until UVI 12. The hair along the right side of the figure is partitioned into three horizontal sections -- red, purple, and blue in ascending order. Each section becomes fully saturated at a different UVI. The bottom turns fully red at UVI 3, the middle turns fully purple at UVI 6, and the top turns fully blue at UVI 12. In other words, the saturation change propagates upward with higher UV intensity. For either design, EcoHair returns to its original color after it is removed from the UV source. \figref{hair-real} shows an actual prototype of the monochromatic wig. The change from negligible to complete saturation took roughly 10 seconds to occur.

\subsection{UV Dosimeter EcoPatch}
\subsubsection{Motivation}
The skin is both a protective barrier against UV radiation and the part of the body most susceptible to damage from excessive UV exposure; therefore, on-body UV sensors placed on exposed skin provide the most relevant information for sunburn risk. Taking inspiration from My UV Patch and LogicInk, EcoPatches are conformal stickers embedded with photosensitive chemicals. The sticker form factor encapsulates the chemicals from the skin, preventing possible irritation. Moreover, stickers can be opaque, making their appearance constant across different underlying skin tones. This is useful for both observations with the naked eye and automated measurements with computer vision.

Our EcoHair prototypes show instantaneous UV intensity, but UV-induced skin damage is a cumulative process. With this in mind, EcoPatches provide colorimetric, visual readouts of cumulative UV exposure using custom inks that can undergo irreversible color changes. The sensitivity a person has to UV radiation depends on their skin tone. The Fitzpatrick scale is a dermatological tool developed to numerically classify skin tones based on their response to UV irradiation \cite{Fitzpatrick1975SoleilPeau,Fitzpatrick1988TheIV}. The scale provides estimates of the minimum amount of UV radiation that leads to solar erythema for six different skin tones. Cumulative UV radiation is measured in UVI-hours (UVIH). A person with pale skin (type I) can get sunburned at around 1 UVIH, while a person with dark brown skin (type VI) can withstand 15 UVIH or higher. 

\subsubsection{Fabrication}
At the foundation of EcoPatches is a custom UV-sensitive ink inspired by Araki \etal{} \cite{Araki2017}. The ink consists of a photoacid generator (PAG), a pH indicator dye solution, a basic buffer, and ethanol. Specifically, our formulations use diphenyliodonium chloride (DPIC) for the PAG and sodium hydroxide (NaOH) for the base buffer. Under UV irradiation, PAGs generate hydrogen ions from photon absorption. The hydrogen ions increase the acidity of the solution, causing the pH indicator to change color. This color change is irreversible unless additional base buffer is introduced into the ink to neutralize the hydrogen ions.

The color spectrum of an EcoPatch depends on the type of pH indicator used in the ink. Therefore, a few design criteria should be considered. First, users should be able to discern distinct color changes as cumulative UV exposure increases from 0 to 12 UVIH. Since pH is a logarithmic scale, a basic solution will be more sensitive than an acidic solution to changes in hydrogen ion concentrations. For example, if a solution starts at a pH 9, changing it to pH 6 requires an increase of $ 9.99^{-7} $ M in hydrogen ion concentration ($10^{-6}-10^{-9} = 9.99^{-7}$ M); however, applying the same concentration change at pH 6 only lowers the pH level to 5.96 ($-\log_{10}[{9.99^{-7}+10^{-6}}] = 5.96$). This means that EcoPatch ink changes its color quickly at low UV exposure levels and slowly at higher levels. If precision is needed at lower levels, a pH indicator with a lower dynamic range should be used. Second, inks should be easier to read if they exhibit roughly linear color changes at the most important cumulative UV exposure levels. This means that pH indicators with narrow operation ranges are more suitable. For EcoPatches, we have prototyped with bromothymol blue, which changes from yellow to blue between pH 2 and pH 7, and phenol red, which changes from yellow to red between pH 6 and pH 8. 

Beyond the pH indicator, there are other factors that affect an EcoPatch's responsiveness. Higher concentrations of buffer slow down the rate of pH change by neutralizing hydrogen ions. Therefore, the base buffer concentration can be used to modulate the amount of cumulative UV exposure the ink can experience before it stops noticeably changing color. A suitable printing medium is also critical since it should not react with the UV-sensitive ink. It is therefore critical that the medium has a neutral pH. Of the substrates we tested, temporary tattoo paper emerged as the best candidate. 

\begin{figure}[!tb]
\begin{center}
\includegraphics[width=\columnwidth]{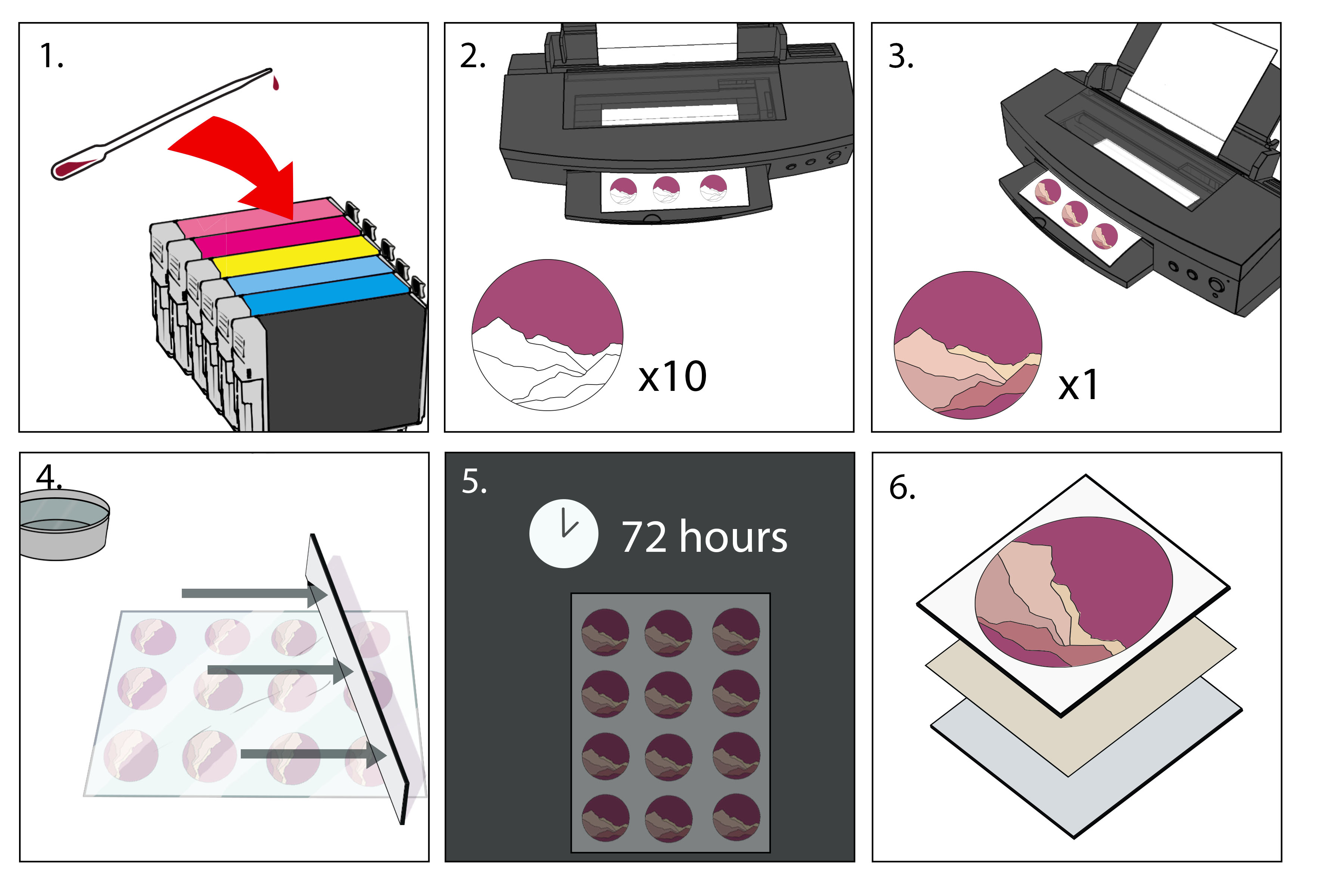}
\end{center}
\caption{The fabrication process for creating a sheet of EcoPatches: (1) UV-sensitive ink is loaded into printer cartridges, (2) The UV-sensitive region of the design is printed onto temporary tattoo paper 10 times, (3) The color reference regions are printed with regular inks, (4) Archival resin is applied on top of the sheet with a squeegee, (5) The sheet is cured for 72 hours in a dark space, and (6) Adhesive films are applied to the sheet.}
\label{fig:patch-process}
\end{figure}

The steps for making a sheet of EcoPatches are illustrated in \figref{patch-process}. First, UV-sensitive inks are filled into a set of regular inkjet printer cartridges. Those cartridges are used to print the UV-sensitive regions of the design on temporary tattoo paper. The UV-sensitive ink is printed through 10 runs on the same paper to ensure enough UV-sensitive ink has been deposited. After the UV-sensitive region has been printed, the UV-sensitive ink cartridges are taken out and replaced with a set of regular printer cartridges. The same printed sheet of patches is loaded back into the printer and a new round of printing adds the color reference region. The sheet of EcoPatches is dried in a dark, ventilated space for 5 minutes and then overlaid with a thin layer of archival grade resin using a squeegee. The resin provides waterproofing, chemical encapsulation, and UV absorption. As a UV absorber, the resin decreases UV transmittance to the ink and enables EcoPatches to measure cumulative exposure beyond 3 UVIH. Thus, applying the right thickness of resin is critical for getting the correct dynamic range. Once the resin has been applied, they are left overnight in a dark, isolated compartment for the resin to cure. A double-sided adhesive film is then applied to the back of the patches. Finally, a skin-safe, single-sided adhesive film is attached to the double-sided adhesive. \figref{patch-process}b shows the full stack of layers in a completed EcoPatch. \tabref{patch-cost} shows the list of materials needed to create a sheet of EcoPatches and their cost.

\begin{table}[!tb]
\centering
\begin{tabularx}{\columnwidth}{|X|X|}
\hline
\textbf{Item} & \textbf{Price (USD)} \\ \hline
    UV-sensitive ink & \$0.50 \\ \hline
    Standard printer ink & \$2.75 \\ \hline
    Temporary tattoo paper & \$3.50 \\ \hline
    Archival grade resin\urllink{https://www.artresin.com/products/8-oz-mini-kit} & \$0.25 \\ \hline
    Double-sided adhesive\urllink{https://www.nitto.com/us/en/products/group/double/037/} & \$0.75 \\ \hline
    Skin-safe adhesive \footnote{The adhesive used for this work is not available to the public yet} & \$0.75 \\
    \hline\hline
\textbf{Total} & \$8.50 \\ \hline
\end{tabularx}
\caption{The bill of materials for creating one sheet of EcoPatches.}
\label{tab:patch-cost}
\end{table}

\subsubsection{Potential Designs}
\begin{figure}[!tb]
\begin{center}
\includegraphics[width=\columnwidth]{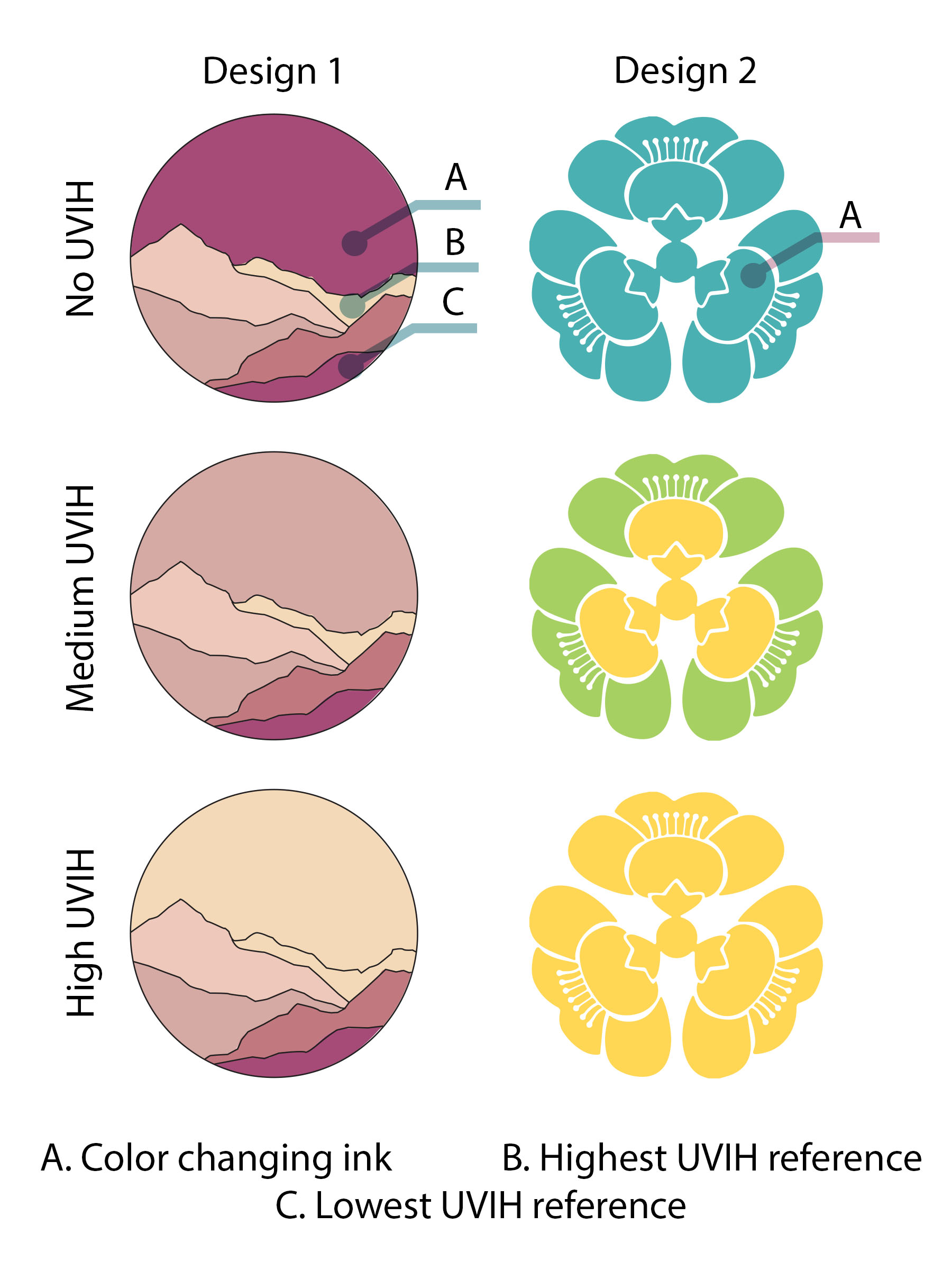}
\end{center}
\caption{EcoPatches can convey quantitative measurements of environmental hazards in different ways. (left) The landscape design includes a reference color scale that the user can use as a point of comparison. (right) The petals in the lotus design start to change their color at different UV exposure levels.}
\label{fig:patch-design}
\end{figure}

As with the EcoHair, we devised two different ways in which EcoPatches could exhibit quantitative measurements; they are shown in \figref{patch-design}. On the left, the patch is divided into a UV-sensitive region (sky) and a color reference region (hills). The UV-sensitive region goes through a continuous color transition from purple to yellow as it receives more UV radiation. The color reference regions are permanent and constant. Each hill represents the sky color associated with a particular level of cumulative exposure. Thus, the user can estimate cumulative UV exposure by comparing the sky color to color references. The reference colors are determined by exposing the printed UV-sensitive ink to sunlight, taking photographs of the ink with a color-calibrating setup (\eg{} a Macbeth Color Checker) when it reaches a desired level for the references, and picking the calibrated color from that image. 

In the design shown on the right of \figref{patch-design}, the patch is divided into three regions with different UV sensitivities. All regions start as turquoise and eventually change to yellow. However, this color transition occurs at a different rate for each region. The innermost region saturates at 3 UVIH, the middle region saturates at 8 UVIH, the outermost region saturates at 12 UVIH. In other words, the lotus propogates a color change from the inside out. The user can estimate their cumulative UV exposure from the portion of the lotus that has turned yellow. 

\begin{figure*}[!tb]
\begin{center}
\includegraphics[width=\textwidth]{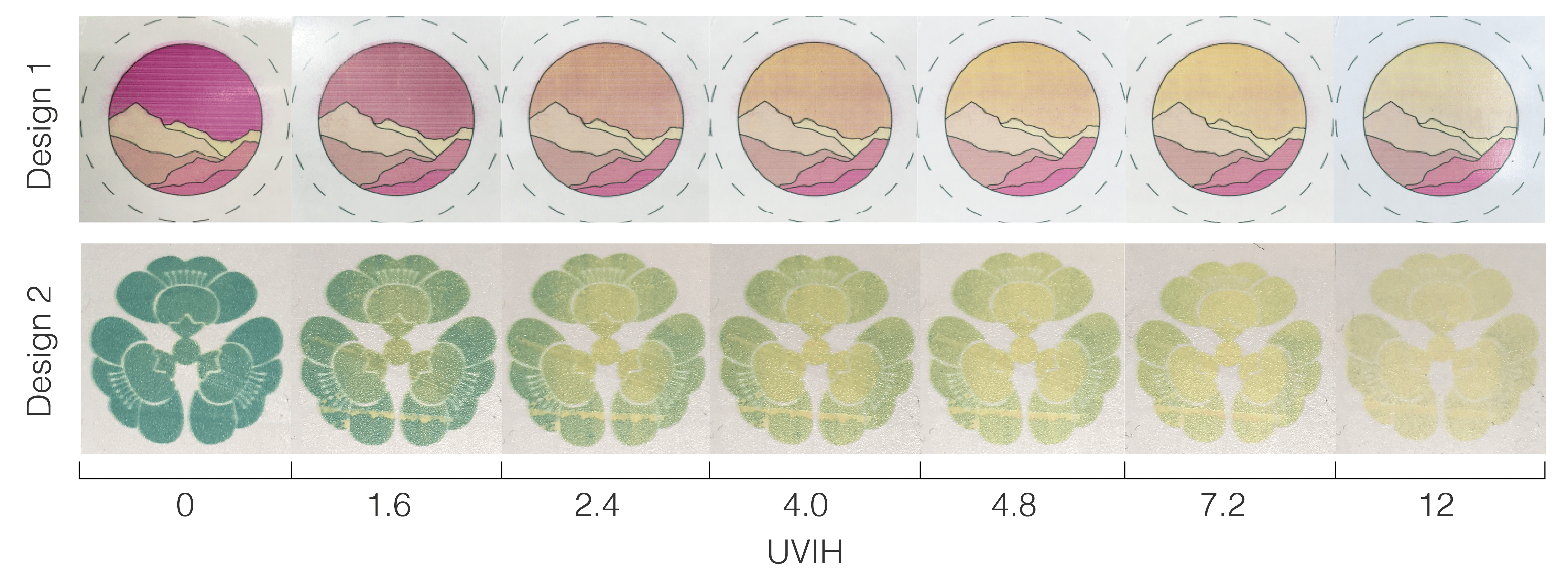}
\end{center}
\caption{From left to right, the photographs show two different EcoPatch designs that have experienced increasing amounts of UV exposure. (top) An EcoPatch with a reference scale made with phenol red (bottom) An EcoPatch without a reference scale made with bromothymol blue.}
\label{fig:patch-real}
\end{figure*}

\figref{patch-real} shows actual photographs of the two different designs over different UVIH measurements. Note that the top patch with the hills was made with phenol red, while the bottom lotus patch was made with bromothymol blue.
\vspace{-0.1in}

\section{Focus Groups}
\label{sec:focusgroup}
We held two focus groups to gauge interest in \projectname{}.
the first focus group was conducted at a technology company and consisted of 13 females and 1 male, aged 20-50 years old. 
The second focus group was conducted at a university and consisted of students with interests in human-computer interaction and design.
The group had 4 females and 1 male, also aged 20-50 years old. 

During each focus group, the interviewers presented the EcoHair and EcoPatch prototypes and described example usage scenarios. Participants were prompted to provide feedback through questionnaires and a discussion session after each prototype was showcased. 

\subsection{Aesthetics That Incentivize Exposure}
After being shown an EcoPatch, one participant was reminded of a product called Mr.~Yuk stickers\urllink{http://www.chp.edu/injury-prevention/teachers-and-parents/poison-center/mr-yuk} (P15, F).
Mr.~Yuk stickers were designed to warn children about toxic substances.
The stickers' creators conducted numerous studies to identify the designs that would be the most unappealing to children so the stickers would act as a deterrent \cite{Fisher1973}.
They converged on a fluorescent green face with a sour expression to warn children about toxic substances, but later research found that the final design may have actually attracted children to harmful substances \cite{Fergusson1982, Vernberg1984}.

\projectname{} accessories are different from Mr.~Yuk stickers because they are dynamic; nonetheless, the story of Mr.~Yuk stickers provides an interesting lesson.
Some people may prefer a \projectname{} accessory's baseline appearance, while others may prefer its appearance after it is exposed to an environmental hazard.
Because of this, an accessory's aesthetic range can incentivize or disincentivize exposure.
Participants felt that the EcoHair incentivized exposure, with some speculating that they ``would (go outside) for fun just to have it change color'' (P15, F).
Although UV exposure is practically unavoidable and harmless in standard doses, participants imagined scenarios where they would be motivated to seek otherwise avoidable hazards.
After being told that chemical sensors could be made for hazards like carbon monoxide, one participant joked that she would wave an EcoPatch behind a car's exhaust just to activate a color change (P13, F).
Future \projectname{} designs should have appealing aesthetics that encourage continuous use, but care must be taken to avoid incentivizing unnecessary exposure.

\subsection{Designing for Aesthetic Transitions}
Participants asked many questions about the spectrum of colors the different sensors could exhibit because of the implications it had on the accessory's aesthetics.
Some people wanted an EcoPatch's design to be aesthetically pleasing regardless of the measurement it was trying to display.
Making an aesthetically pleasing design with a region that can be any shade of green is easier than making one for all colors of the rainbow.
From a technical standpoint, however, more extreme color changes are preferred since they improve an accessory's readability and precision.
This raises a tension between aesthetics and accuracy that motivates collaboration between chemical engineers and artists early in the design process.
As a remedy to this tension, one participant suggested EcoPatch designs that change in color opacity, such as a design that transitions from empty black outlines to a color-filled design at full exposure.

\subsection{Permanence of Aesthetic Changes}
Our EcoPatch prototype encodes a cumulative measurement, thus implying a monotonic change over time.
As such, our EcoPatches are designed to be replaced on a daily basis.
However, some participants preferred reusable EcoPatches, thereby limiting material waste and repeated purchases.
Creating a cumulative sensor that reacts in one direction when it is exposed to an environmental hazard and in the other when it is not exposed would require two separate chemical reactions, making the fabrication process much more difficult.
A reversible cumulative reaction would also complicate the interpretation of the measurement.

Nevertheless, an opportunity for future research lies in chemical reactions that can be manually reversed by the user.
For instance, participants envisioned EcoPatches that could be reversed by placing them under running water or inside a refrigerator. 
Alternatively, an instantaneous measurement could be included alongside the cumulative one so that the user can benefit from their accessory after the cumulative region has been saturated.

\subsection{Visibility of Aesthetic Changes}
Both focus groups included discussions about who consumes the information conveyed by the EcoHair.
Because most hair lies above or behind the head, a person's peers are more likely to see the EcoHair change colors than the wearer themselves.
This leads a person to be unaware about their own body, which can be startling.
As one participant exclaimed, ``What freaks me out is that everyone else will see first what's going on with me, and then myself'' (P2, F).
Although \projectname{} accessories are primarily framed as ambient sensors, EcoHair actually acts as a publicly displayed sensor in many scenarios.

The visibility of a \projectname{} accessory is a function of both its form factor and aesthetic design, so designers have complete flexibility in how they create future sensors.
In some cases, a designer may actually want to create an ostentatious display in order to incite a reaction about fashion or the environment.
If a designer tones down an accessory's aesthetics too much, the accessory may become so discreet that the wearer forgets about it; this sentiment was occasionally expressed by focus group participant regarding the EcoPatch.
Therefore, designers should be constantly aware of who consumes the data from their accessory and how often the data is consumed.

\section{Future Work and Limitations}
\label{sec:future}

\subsection{New Form Factors}
The convenience of \projectname{} sensors makes them well-suited for people who travel to unfamiliar places. Instead of having sensors that double as decorative accessories, makers can embed \projectname{} inks into functional fashion items that are already worn on a daily basis (\eg{} shoelaces, watch wristbands). It should be noted that the inks' chemical properties are substrate-dependent, so makers may need to recalibrate measurement benchmarks for their given substrate. 

Participants from the focus group expressed interest in form factors that could be easily hidden. Towards this desire, we have explored false nails and nail polish. People's hands come into contact with innumerable foreign substances throughout the day, making them ideal for chemical sensing. For instance, because fingers often come into contact with tap water via hand washing, false nails could be useful for detecting waterborne toxins and contaminants. Examples of functional nail accessories include nail polish that changes color when it comes into contact with date-rape drugs like Rohypnol, Xanax, and GHB \cite{murray2012molecularly}. Jewelry, in particular rings and necklaces, is another form factor that can be easily concealed. Moreover, decorative pieces like necklaces have the option of being displayed or hidden depending on user preferences. 

Another potential form factor is sensors that can simultaneously block and measure environmental hazards. One suggestion that arose during the focus groups was a mouth mask that blocks and measures air pollution. Surgical and N95 masks are popular in East Asian countries for reducing the spread of airborne disease and exposure to air particulates\cite{Yokota2016}. Due to their popularity, mouth masks come in designs ranging from geometric patterns to cute animal faces, making them a natural fit under \projectname{}.

\subsection{Other Environmental Hazards}
Water contamination is a serious health concern in many parts of the world. Today, at least 2 billion people globally still use a contaminated source for drinking water \cite{WHO2018}. Within the United States, close to 20 million people may be served by water systems that are not properly monitored or maintained for lead levels, and close to 4 million people are served by water systems with lead levels exceeding the EPA action level of 15 parts per billion \cite{Olson2016}. Citizen sensing could raise awareness to serious environmental violations. Flint, Michigan is an example of citizen science effectively exposing an environmental crisis in light of institutional failures to inform the public \cite{AssociatedPress2015FlintWater}. As mentioned earlier, a false nail form factor is useful for convenient liquid measurements, so we are developing false nails that indicate elevated lead concentrations in tap water through color changes. It has been shown that DNA circuits coupled with gold nanoparticles can be used to detect lead concentrations in water and display a colorimetric readout \cite{Liu2003,Wei2008}. The nails we are developing encapsulate this DNA-based molecular circuit in a PVC-like matrix. 

Air pollution is another environmental hazard that causes adverse health effects. One of the gases that make up urban smog is carbon monoxide (CO). Exposure to CO can lead to cardiovascular and neurological impairments. Although the effects of indoor CO poisoning are widely known, new evidence suggests that chronic exposure to low concentrations of CO, such as those found in vehicle exhausts, can lead to long-term, detrimental health effects \cite{WHO2017}. Although CO is not a major component of gaseous air pollutants, micro-scale CO tracking may be useful as a proxy for determining air quality in urban centers \cite{Alm1999}. EcoPatches for CO are chemically based sensors designed to detect CO in the air and provide a colorimetric readout indicative of CO levels. Previous research has demonstrated powders that change color when exposed to CO \cite{Moragues2011}. We have experimented with similar powders dissolved in ethanol and used the EcoPatch fabrication protocol to make CO-sensing stickers. Similar procedures could be applied to making EcoPatches for other air pollutants, such as NO$_x$ and SO$_x$ gases. 

\subsection{Integration with Computing}
So far, \projectname{} accessories have been described as standalone analog sensors untethered from batteries or transmitters. However, the fact that \projectname{} sensors transduce invisible chemicals into visible color changes provides opportunities for commodity hardware to measure environmental hazards. People with poor color acuity or colorblindness can benefit from a smartphone app that interprets the colorimetric readout for them. We are developing a companion app which analyzes photos of \projectname{} accessories. For patterned designs like the landscape EcoPatch, the app can locate the design using feature matching and segment it using homographic transformations. The app can then interpolate the UV-sensitive region's color between the references in a color space (\eg{} RGB, HSV) to estimate a measurement. 

Integrating \projectname{} with an app can also provide similar benefits to other self-tracking apps like step counters or food diaries.
By storing measurements over time, a person can examine a personal history of their data.
By uploading their measurements to the cloud, users can connect to a community of \projectname{} citizen scientists. Comparing geospatially arranged data would allow people to identify neighborhoods or districts where environmental hazards are particularly elevated.

\subsection{Gender Imbalance}
All but two of our participants identified as female, but there was no intention of focusing on a specific gender during recruitment. Part of this could have been due to the explicit branding of \projectname{} as fashion accessories during recruitment. Advertising form factors and designs that appeal to different genders or gender-neutral form factors and designs that appeal to all could balance recruitment in the future.
\section{Conclusion}
\label{sec:conclusion}

The impact environmental hazards have on our personal health is difficult to grasp partly because the hazards themselves are invisible.
\projectname{} aims to make these hazards more apparent using fashion as an accessible medium.
Even for people who are not particularly motivated to learn about their environment, \projectname{} provides another outlet for self-expression via dynamic clothing and accessories.
We described two different chemical sensors that respond to UV radiation and then outlined how they can be integrated into fashion accessories using fabrication processes that can be performed at home.
Our focus groups raised questions about the role aesthetics play in encouraging usage, conveying information, and benefiting the wearer's peers. 
Advances in programmable materials research promise to fundamentally alter how we perceive and interact with the physical world. 
We hope this research helps to illuminate the possibilities of a future where programmable materials are used to make everyday objects more dynamic and informative. 
\section{Acknowledgments}
\label{sec:acknowledgments}
We thank Hsin-Liu (Cindy) Kao for her formative work on powder-based chemical sensors, which served as inspiration for the colorimetric sensors and interaction modalities described in this work. 
We thank Sarah-Grace Lingenbrink, Shine Lingenbrink, and Wende Copfer for their help in creating a companion video for this project. 
We also thank Sarah-Grace for her help with integrating the EcoHair chemicals into salon-ready products.

\balance{}

\bibliographystyle{ACM-Reference-Format}
\bibliography{ms}

\end{document}